# High-pressure optical floating-zone growth of Li$_2$FeSiO$_4$ single crystals

Waldemar Hergett[a,*], Christoph Neef[a,g], Hubert Wadepohl[b], Hans-Peter Meyer[c], Mahmoud M. Abdel-Hafiez[a,d], Clemens Ritter[f], Elisa Thauer[a], Rüdiger Klingeler[a,e]

[a]Kirchhoff Institute of Physics, Heidelberg University, 69120 Heidelberg, Germany
[b]Institute of Inorganic Chemistry, Heidelberg University, 69120 Heidelberg, Germany
[c]Institute of Earth Sciences, Heidelberg University, 69120 Heidelberg, Germany
[d]Physics Department, Faculty of Science, Fayoum University, 63514 Fayoum, Egypt
[e]Centre for Advanced Materials (CAM), Heidelberg University, 69120 Heidelberg, Germany
[f]Institut Laue-Langevin, 38042 Grenoble, France
[g]Fraunhofer Institute for Systems and Innovation Research ISI, 76139 Karlsruhe, Germany

**Abstract**

We report the growth of mm-sized *Pmnb*-Li$_2$FeSiO$_4$ single crystals by means of the optical floating-zone method at high argon pressure and describe the conditions required for a stable growth process. The crystal structure is determined and refined by single-crystal X-ray diffraction. The lattice constants amount to *a* = 6.27837(3) Å, *b* = 10.62901(6) Å and *c* = 5.03099(3) Å at 100 K. In addition, we present high-resolution neutron powder diffraction data that suggest that the slight Li – Fe site exchange seems to be intrinsic to this material. High quality of the crystal is confirmed by very sharp anomalies in the static magnetic susceptibility and in the specific heat associated with the onset of long-range antiferromagnetic order at $T_\mathrm{N}$ = 17.0(5) K and pronounced magnetic anisotropy for the three crystallographic axes. Furthermore, magnetic susceptibility excludes the presence of sizable amounts of magnetic impurity phases.

1. **Introduction**

The orthosilicate Li$_2$FeSiO$_4$ has been attracting attention as a potential new generation cathode material for lithium-ion batteries as it offers a low-cost and environmentally friendly compound for large-scale energy storage applications. In particular, silicate cathode materials represented by Li$_2$FeSiO$_4$ provide high thermal stability, good operation voltage and promise high capacity [1, 2]. The strong Si – O bond possesses high chemical resistance towards electrolytes. Optimization of the electrochemical performance and the rich temperature-dependent polymorphism of Li$_2$FeSiO$_4$ have hence been a subject of intensive research (e.g. [3, 4, 5, 6, 7, 8]). However, while there is a variety of reports on the structural, morphological and electrochemical properties of Li$_2$FeSiO$_4$, due to the lack of single crystals neither single-crystal structure refinement of *Pmnb*-Li$_2$FeSiO$_4$ nor any studies of anisotropic electronic properties relevant for application in lithium-ion batteries have been reported yet.

In order to investigate in detail structural effects, magnetic ordering, magnetocrystalline anisotropy and other intrinsic properties, comprehensive studies on appropriately large and high-quality single crystals are mandatory. Here, we report the growth of Li$_2$FeSiO$_4$ single crystals featuring the *Pmnb* phase by means of the optical traveling floating-zone technique at high argon pressure of 30 bar. We present full structure refinement, which was solved by single-crystal X-ray diffraction. In addition, high-resolution neutron powder diffraction is applied to refine Li positions further in Li$_2$FeSiO$_4$ as site exchange between Li and Fe is supposed to be relevant for Li-ion mobility [9]. Finally, thermodynamic studies of the static magnetization and the specific heat not only confirm the high quality of the grown crystals but also enable investigating magnetic anisotropy and the evolution of long-range antiferromagnetic order at $T_\mathrm{N}$.

2. **Experimental**

The crystal growth was carried out in a high-pressure floating-zone (FZ) furnace (HKZ, SciDre) [10]. Polycrystalline Li$_2$FeSiO$_4$ starting materials used for the FZ process have been synthesized by means of a conventional one-step solid-state reaction yielding the $Pmn2_1$-polymorph. Ferrous oxalate dihydrate $FeC_2O_4·2H_2O$ with a 5 % excess above stoichiometric requirements, lithium carbonate and silicon dioxide were mixed in a planetary high-energy ball mill with acetone, dried and heated at 370°C for 12h. Subsequently, the product was reground, pelletized and sintered at 800 °C for 6 h using a heat-up ramp of 300 °C/h. The tube furnace was operated at 100 mbar with a constant Ar flux of 250 standard cubic centimeters per minute and pressurized to 1400 mbar under static atmosphere after heating up to 430 °C in order to mitigate Li volatility. Reground powder was compacted into feed rods with diameters of 6 mm and typical lengths of 70 to 110 mm under an isostatic pressure of 60 MPa. The feed rods thus produced were used without sintering for all crystal growth experiments. For high-resolution neutron diffraction, polycrystalline $Pmnb$-Li$_2$FeSiO$_4$ was synthesized by a solid-state reaction between stoichiometric amounts of $Li_2SiO_3$, $Fe_2O_3$ and Fe. The precursors were ball-milled with acetone, dried and transferred to the tube furnace. After heating the mixture to 800 °C for 12 h the temperature was raised to 950 °C for another 16 h. Subsequently, the product was quenched to room temperature by immersing the crucible in a water bath. Synthesis conditions concerning heat-up ramps and tube atmosphere were chosen as described above.

The polycrystalline samples as well as the ground single crystals obtained after the FZ growth were studied by means of powder X-ray diffraction using Cu-K$\alpha_{1,2}$ radiation on a Bruker D8 Advance ECO diffractometer equipped with an SSD-160 line-detector in Bragg-Brentano geometry. Data have been collected at room temperature in the 2θ range of 10 ° to 70 ° with 0.02 ° step-size and 2.4 s integration time. The synthesized feed rods exhibit the $Pmn2_1$ phase. Powder neutron diffraction data were collected on the D2B diffractometer at the Institut Laue-Langevin (ILL), France, using a wavelength of 1.051 Å corresponding to the (557) reflecting planes of a germanium monochromator. For Rietveld refinement of both neutron and X-ray powder diffraction profiles the FullProf 2.0 software was used [11]. SEM-EDX microanalysis of the resulting crystals was conducted by means of a Leo 440 scanning electron microscope equipped with an Inca X-Max 80 detector (Oxford Instruments). The acceleration voltage was 20 kV, the working distance was 25 mm, and the counting time was 100 s (lifetime) at about 10,000 cps. Single-crystal X-ray studies were performed at 100 K using an Agilent Technologies Supernova-E CCD 4-circle diffractometer (Mo-K$\alpha$ radiation λ = 0.71073 Å, microfocus X-ray tube, multilayer mirror optics).

X-Ray Laue diffraction in back scattering geometry was used to orient the single crystals, which were then cut to cuboids with respect to the crystallographic main directions. Laue diffraction was done on a high-resolution X-Ray Laue camera (Photonic Science). The surfaces of the oriented samples and of the cut boules were inspected by polarized light optical microscopy. Magnetization was measured using a superconducting quantum interference device (SQUID) magnetometer (Quantum Design MPMS-XL5). Specific heat measurements were carried out in a Quantum Design PPMS using the relaxation method.

3. **Crystal growth**

The crystal growth employing polycrystalline starting materials was carried out in a high-purity argon atmosphere, at a pressure of 30 bar and a flow rate of 0.02 l/min. The feed and seed rods were rotated in opposite directions for homogenization of the molten zone at a rate of 21 and 17 rpm, respectively.

Successful growth of crystals was performed at elevated Ar pressure. Indeed, it is a common feature of Li-containing systems that pressure mitigates volatilization of lithium oxide during crystal growth. This is e.g. evidenced by the amount of material condensing in the growth chamber, which provides one of the various guiding principles optimizing our growth processes [12, 16]. The actual chosen pressure of 30 bar results from our preliminary studies at various pressures.

Fig. 1 shows the temperature distribution recorded along the rods' axis after formation of the molten zone. The curve exhibits a bell-like shape with a broad central region and a plateau-like shoulder at around $z$ = 15 mm. The shoulder is associated with the liquid-solid coexistence region [12] which reflects the melting temperature $T_m$ = (1240 ± 5) °C. For the growth process, the molten zone was stabilized regarding shape and volume at a temperature of 1310 °C. Sharp focusing of the light, high Ar gas pressure and the material's properties (e.g. thermal conductivity) yielded steep temperature gradients of up to 150 °C/mm along the growth direction [10, 13, 14, 15, 16]. A slightly convex growth front is revealed by the longitudinal cross-section of the frozen zone (Fig. 1b).

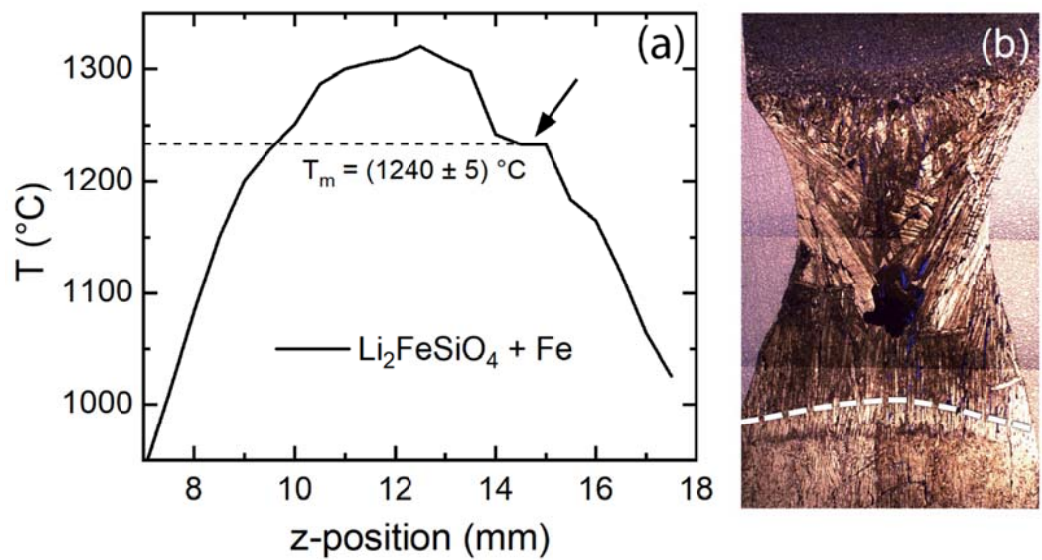

Figure 1: (a) Temperature profile recorded along the vertical axis of the feed rod in the region of the molten zone. The shoulder indicated by the arrow is associated with melt-solid coexistence and enables reading off the melting temperature $T_m$. (b) Optical polarized light image of a longitudinal cross-section of the frozen zone. The white dashed line marks the zone-crystal interface.

## 4. Characterization

The XRD pattern of the powdered crystal indicates phase pure material. Rietveld refinement of the experimental data in Fig. 2 shows that all observed diffraction peaks are associated with the *Pmnb* phase of $Li_2FeSiO_4$. We note that the diffraction pattern displays enhanced peak intensity for (020), indicating preferred orientation. This factor was included in the Rietveld refinement of the data. We attribute this to the anisotropic tendency to agglomeration, favoring oriented aggregation of the platelet-shaped crystallites (see the inset in Fig. 2) originating from anisotropic cleavage properties (see, e.g., Refs. [17, 18]). At room temperature, the lattice parameters were determined as $a$ = 6.280(5) Å, $b$ = 10.652(9) Å and $c$ = 5.037(3) Å.

The *Pmnb* structure matches with the expected high-temperature phase of $Li_2FeSiO_4$ [3]. We note, that by choosing stoichiometric amounts of $FeC_2O_4·2H_2O$ for the initial synthesis step, about 10-15 % of the $Li_2SiO_3$ impurity phase was found in the as-grown boule. In addition, stoichiometric starting material yields detrimental effects on the crystal morphology in terms of formation of the $P12_1/n1$-polymorph. Our results show that excess of $FeC_2O_4·2H_2O$ significantly improves the phase purity of the grown crystals. We hence have used 5 % of $FeC_2O_4·2H_2O$ above stoichiometric proportions for the preparation of the starting materials, which results in the presence of $α$-Fe in the feed rods in agreement with [19]. Using the $P12_1/n1$-polymorph as the starting material does not yield phase pure single crystals. In principle, a structural phase transition *Pmnb* → $P12_1/n1$ is also feasible at a slow cooling rate since, according to thermodynamic calculations, the *Pmnb*-polymorph is energetically less favorable so that metastable behaviour could be envisaged [20]. However, in our experiments we did not observe any influence of the growth rate, which was varied between 1.5 and 30 mm/h, on the polymorphism of the obtained crystals in contrast to the optimization of the synthesis conditions. We conclude that different cooling rates are not a governing parameter.

Elemental composition of the grown crystals was determined by means of energy-dispersive X-ray spectroscopy (EDX). The EDX data were acquired as well from several point scans as from area scans (mapping) on polished surfaces. A typical grain extracted from the as-grown boule pulled at a rate of 10 mm/h is seen in Fig. 3. The formation of two well distinct phases can be easily identified by the different contrast in the backscattered electron (BSE) image. The dominant phase indicated by the light gray regions corresponds to the main phase of $Li_2FeSiO_4$.

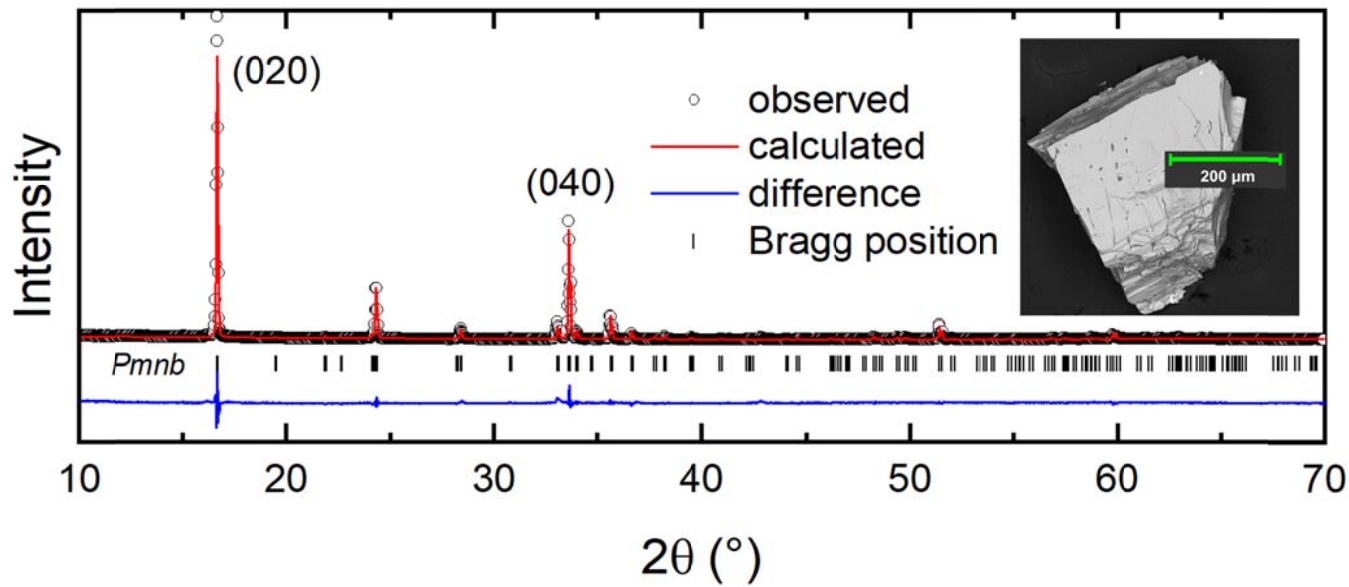

Figure 2: XRD pattern collected from the powdered single crystal (open circles). The (red) line shows the simulated pattern, the blue one below displays the difference of observed and calculated data. Vertical ticks designate the Bragg positions. Inset: SEM image of a single-crystalline grain with prominent cleavage.

In addition, there are small regions of bright inclusions associated with iron oxide as indicated by the EDX data (see below). These inclusions show branching morphology with up to several hundred micrometer-sized dendrites. EDX point scans at positions c1 and c2 (see Fig. 3) confirm a Si:Fe ratio that is very close to that of the nominal stoichiometry of $Li_2FeSiO_4$. In the dendrites (positions c3 and c4), our data verify absence of Si while the amount of oxygen cannot be quantified with the required significance by EDX microanalysis. Dark gray spots in the rim region are presumably caused by pores and microbubbles. Area scans (marked in Fig. 3 by the rectangles c5 and c6) do not show any compositional modulation compared to the main phase measured in the center of the grain, i.e., at the positions c1 and c2. The EDX results are summarized in Tab. 1. The stated errors represent standard deviations of the averaged measurements. For our analysis, we have balanced the positive charges of the cationic sites by an appropriate quantity of oxygen assuming a nominal oxidation state of +2 for Fe. Since no reference standards with a known Si:Fe ratio were available and due to the high atomic weight difference of these elements, the quantitative reliability of the EDX data is limited, but still sufficient to estimate the compositions and thereby identify phases.

Table 1: Results of the EDX analysis at the measured points and areas indicated in Fig. 3.

|  | main phase (c1, c2, c5, c6) | $FeO_x$ impurities (c3, c4) |
|---|---|---|
| $SiO_2$ (mol%) | 0.52(1) | 0.003(1) |
| FeO (mol%) | 0.48(1) | 0.997(5) |

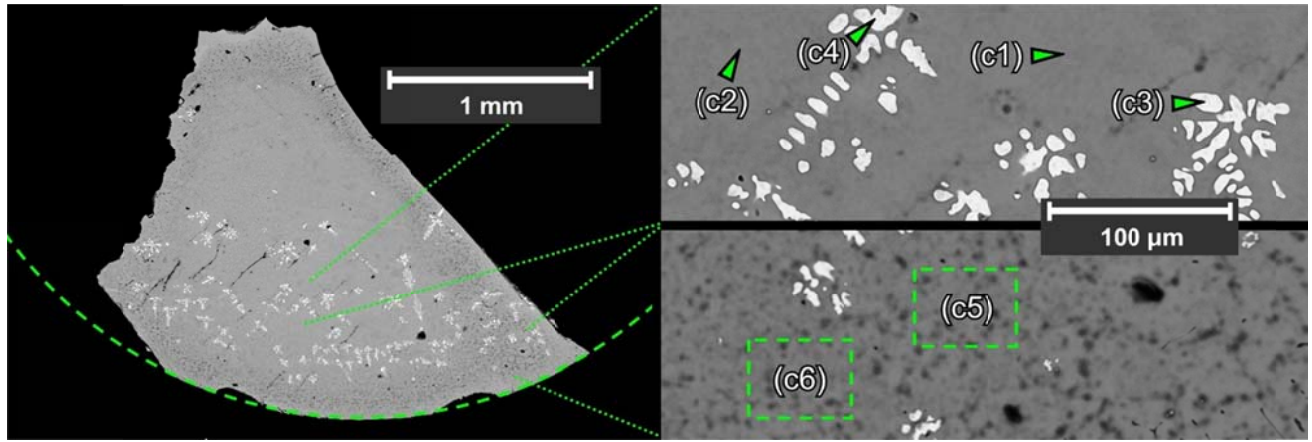

Figure 3: BSE image of a $Li_2FeSiO_4$ grain with local enlarged views of the dendritic $FeO_x$ inclusions. EDX measurement points and areas are marked and labeled c1 to c6.

In Fig. 4 EDX spectra of the $Li_2FeSiO_4$ main phase (a) and the dendritic inclusions are displayed. A small peak at ≈ 0.28 keV is caused by carbon coating of the sample. Peaks associated with silicon are not observed in the $FeO_x$-phase. For semi-quantitative processing of this phase, simulated spectra for $FeO$, $Fe_2O_3$ and $Fe_3O_4$ were fitted and normalized to the maximum value of the Fe-K$\alpha$ peak. All fitted curves show significant deviation from the experimental Fe-L$\alpha_1$ and Fe-L$\beta_1$ peak intensities. The best correspondence for the O-K$\alpha_1$ peak is obtained by assuming an iron(II) oxide phase. Analysis of polycrystalline $Li_2FeSiO_4$ powders in [3] and [21] also reveals traces of $Fe_{1-\delta}O$ in a few percent range. In the XRD pattern (Fig. 2), there is no signature of any iron oxide impurity phases. We conclude that the amount of dendritic material is below the detection limit. This assumption is corroborated by the magnetic susceptibility data presented below which suggest ≤ 0.5 % of $Fe_{1-\delta}O$, < 0.03 % of $\alpha$-$Fe_2O_3$, and no $Fe_3O_4$ in the sample.

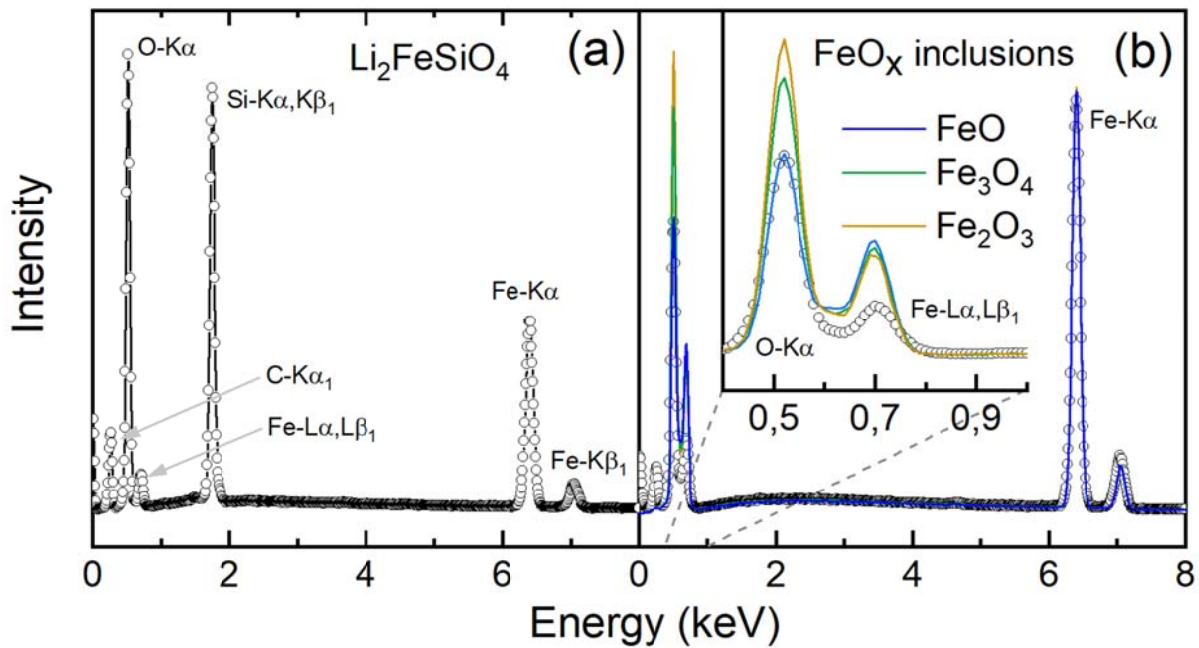

Figure 4: EDX spectra collected from the main (a) and the dendritic (b) phase. Open circles denote experimental values. Simulated spectra for peak fitting analysis of the iron oxides are indicated by solid colored curves. The best fit is obtained for an assumed iron to oxygen ratio of ≈ 1:1.

Several large crystalline grains were oriented by means of X-ray Laue back-reflection with respect to the crystallographic main directions and subsequently cut to oriented cuboids. $Li_2FeSiO_4$ forms air- and moisture-stable, translucent dark brown crystals. Fig. 5 shows an exemplary single crystal of several $mm^3$ in volume. High crystallinity of this sample is illustrated by the Laue images (Fig. 6) taken perpendicular to the (100)- and the (010)-face. The growth direction is parallel to the *bc*-plane, i.e., perpendicular to the *a*-axis. The Laue images exhibit clear diffraction spots and provide already strong indication of the orthorhombic Laue group *mmm*. The overlaid simulated pattern in Fig. 6 confirms the orthorhombic *Pmnb*-symmetry found in the powder XRD pattern. The comparison of Laue patterns obtained from opposite faces of the cuboids assert its uniform crystallinity and orientation. However, it can be observed that a few Laue reflections, e.g. (110), split into separate spots. Since the scattering volume is limited to a near-surface region due to the limited penetration depth of the X-ray beam, we attribute this splitting to dislocated subgrains at the surface [22].

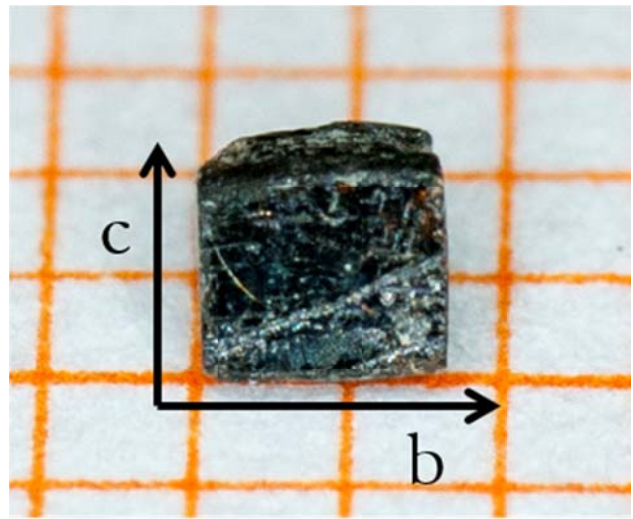

Figure 5: Picture of an oriented $Li_2FeSiO_4$ single crystal on graph paper with a 1 mm grid.

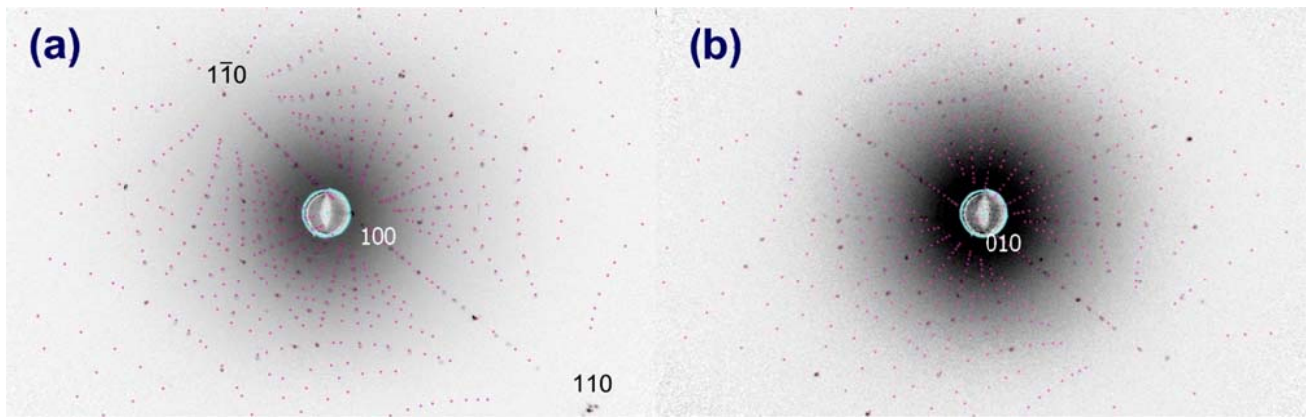

Figure 6: Experimental (black spots) and simulated (purple spots) Laue patterns with the incident beam perpendicular to (a) the (100)- and (b) the (010)-face.

*4.1. Single-crystal structure determination*

Successful growth of single crystals enables solving the crystal structure of *Pmnb*-$Li_2FeSiO_4$. A suitable splinter-like crystal with dimensions of 0.141 × 0.116 × 0.048 $mm^3$ was selected for collection of intensity data. Detector frames (typically ω-, occasionally φ-scans, scan width 0.5 °) were integrated by profile fitting [23, 24]. Data were corrected for air and detector absorption, Lorentz and polarization effects [24] and scaled essentially by application of appropriate spherical harmonic functions [24, 25, 26]. Absorption by the crystal was treated numerically (Gaussian grid) [26, 27]. An illumination correction was performed as part of the numerical absorption correction [26]. The structure was solved by the charge flip procedure [28] and refined by full-matrix least squares methods based on $F^2$ against all unique reflections [29]. All atoms were given anisotropic displacement parameters. For reasons discussed elsewhere [10] ionic scattering factors were employed for $Li^+$ [30], $Fe^{2+}$ [30] and $O^{2-}$ [31]. Refinement of secondary extinction according to the empirical Larson formula [32] resulted in a negligible correction which was not applied to the data.

The final structural model was obtained assuming fully occupied individual sites and no cation mixing. Refinement data including the cell parameters are summarized in Tab. 2, atomic coordinates and equivalent isotropic displacement parameters are listed in Tab. 3, and selected bond lengths and angles are shown in Tab. 4. The single-crystal X-ray diffraction study shows that the sample is of good crystallinity and there is no evidence for the presence of any sizeable crystalline impurity phases consistent with the powder XRD analysis.

The structure refinement (Tab. 2) confirms the *Pmnb* space group and improves the accuracy of atomic coordinates, bond lengths and angles previously obtained from powder XRD in Ref. [21]. The structure of *Pmnb*-$Li_2FeSiO_4$ comprises pseudo-hexagonally close-packed oxygen arrays stacked along the *c* direction. One half of the tetrahedral interstitials in this oxygen framework are occupied by Li, Fe, and Si so that along the *b* direction layers of vertex-sharing $LiO_4$ tetrahedra alternate with layers that contain $FeO_4$ and $SiO_4$ tetrahedra without face-sharing (see Fig. 7a). Within each type of layer, $LiO_4$ and $FeO_4/SiO_4$ tetrahedra are arranged in chains propagating along the *a*-axis (Fig. 7b). In $FeO_4$–$SiO_4$ chains, tetrahedra periodically take opposite orientations whereas in $LiO_4$ chains tetrahedra are oriented identically. This alternation necessitates every $FeO_4$ tetrahedron to share two edges with adjacent $LiO_4$ tetrahedra leading to a distortion as a consequence of different ionic radii of $Fe^{2+}$ (92pm) and $Li^+$ (90pm) [34]. On the other hand, $SiO_4$ tetrahedra exhibit nearly perfect symmetry since they do not share edges and Si – O bonds are of strong covalent character. In $FeO_4$, the bond-length distribution varies by ≈ 7 % from 1.9752(6) Å to 2.1090(9) Å (see Tab. 4). Thus, $Fe^{2+}$ cations undergo distinct off-center displacements within tetrahedra. In perfect tetrahedral coordination, $Fe^{2+}$ would adopt a high-spin configuration (*S* = 2) with a Jahn-Teller (JT) active orbitally degenerate $^2E$ electronic state [35]. The observed distortion of the coordinating oxygen tetrahedron lifts orbital degeneracy. Edge-sharing of $LiO_4$ and $FeO_4$ tetrahedral units results in reasonably small interatomic distances of 2.7632(18) Å between $Li^+$ and $Fe^{2+}$ ions. This agrees with the fact that the *Pmnb*-polymorph of $Li_2FeSiO_4$ is thermodynamically not particularly stable [20].

Table 2: Selected crystallographic parameters of Li$_2$FeSiO$_4$.

| | |
|---|---|
| Empirical formula | Li$_2$FeSiO$_4$ |
| Formula weight | 161.81 |
| $T$ (K) | 100(1) |
| Crystal system, space group | Orthorhombic, *Pmnb* (*IT* Nr. 62) |
| Unit cell dimensions[*] | |
| $a$ (Å) | 6.27837(3) |
| $b$ (Å) | 10.62901(6) |
| $c$ (Å) | 5.03099(3) |
| $V$ (Å$^3$) | 335.732(3) |
| $Z$, Calculated density (g cm$^{-3}$) | 4, 3.201 |
| Absorption coefficient for Mo-K$\alpha$ (mm$^{-1}$) | 4.688 |
| Transmission factors: max, min | 0.840, 0.566 |
| Theta range for data collection | 3.8 ° to 45.4 ° |
| Reflections collected / independent | 60175 / 1500 [R$_{int}$ = 0.0321] |
| Completeness to 0.5 Å | 99.9 % |
| Observed reflections [I > 2σ(I)] | 1434 |
| Data / restraints / parameters | 1500 / 0 / 43 |
| Goodness-of-fit on $F^2$ | 1.102 |
| Final $R$ indices [$F_0 > 4\sigma(F_0)$] $R(F)$,$wR(F^2)$ | 0.0243, 0.0719 |
| Final $R$ indices (all data) $R(F)$,$wR(F^2)$ | 0.0254, 0.0727 |

[*]Standard uncertainties of the cell constants are based on the statistical analysis of the refinement against a large subset of the measured reflections only [24] (no systematic error contributions, Type A as defined by IUCr [33]).

Table 3: Fractional atomic coordinates and equivalent isotropic displacement parameters (Å$^2$). U$_{eq}$ is defined as one third of the trace of the orthogonalized U$_{ij}$ tensor.

| Atom | Wyckoff | x | y | z | U$_{eq}$ |
|---|---|---|---|---|---|
| Li | 8d | 0.4947(3) | 0.3313(2) | -0.7072(4) | 0.006(1) |
| Fe | 4c | 0.75 | 0.4185(1) | -0.2995(1) | 0.005(1) |
| Si | 4c | 0.75 | 0.5839(1) | 0.1934(1) | 0.003(1) |
| O1 | 8d | 0.4619(1) | 0.3439(1) | -0.3058(1) | 0.005(1) |
| O2 | 4c | 0.75 | 0.4364(1) | -0.7170(2) | 0.005(1) |
| O3 | 4c | 0.75 | 0.5904(1) | -0.1312(2) | 0.004(1) |

Table 4: Selected bond lengths (Å) and angles (°). Numbers in boldface denote the M – O distances (M = Fe, Si, Li). The other entries in the table represent the bond angles Oi – M – Oj (i, j = 1 to 3).

|    | O1 | O2 | O3 | O1 |
|----|----|----|----|----|
|    | \multicolumn{4}{c}{FeO$_4$} | | | |
| O1 | **1.9752(6)** | 91.17(2) | 111.781(19) | 132.61(4) |
| O2 |  | **2.1090(9)** | 109.69(4) | 91.17(2) |
| O3 |  |  | **2.0137(9)** | 111.781(19) |
| O1 |  |  |  | **1.9752(6)** |
|    | \multicolumn{4}{c}{SiO$_4$} | | | |
| O1 | **1.6366(7)** | 110.81(3) | 108.98(3) | 108.78(5) |
| O2 |  | **1.6314(10)** | 108.45(5) | 110.81(3) |
| O3 |  |  | **1.6347(9)** | 108.98(3) |
| O1 |  |  |  | **1.6366(7)** |
|    | \multicolumn{4}{c}{LiO$_4$} | | | |
| O1 | **2.035(2)** | 115.21(10) | 108.03(9) | 109.25(9) |
| O2 |  | **1.9541(19)** | 113.36(10) | 115.21(10) |
| O3 |  |  | **1.9271(19)** | 114.66(10) |
| O1 |  |  |  | **1.946(2)** |

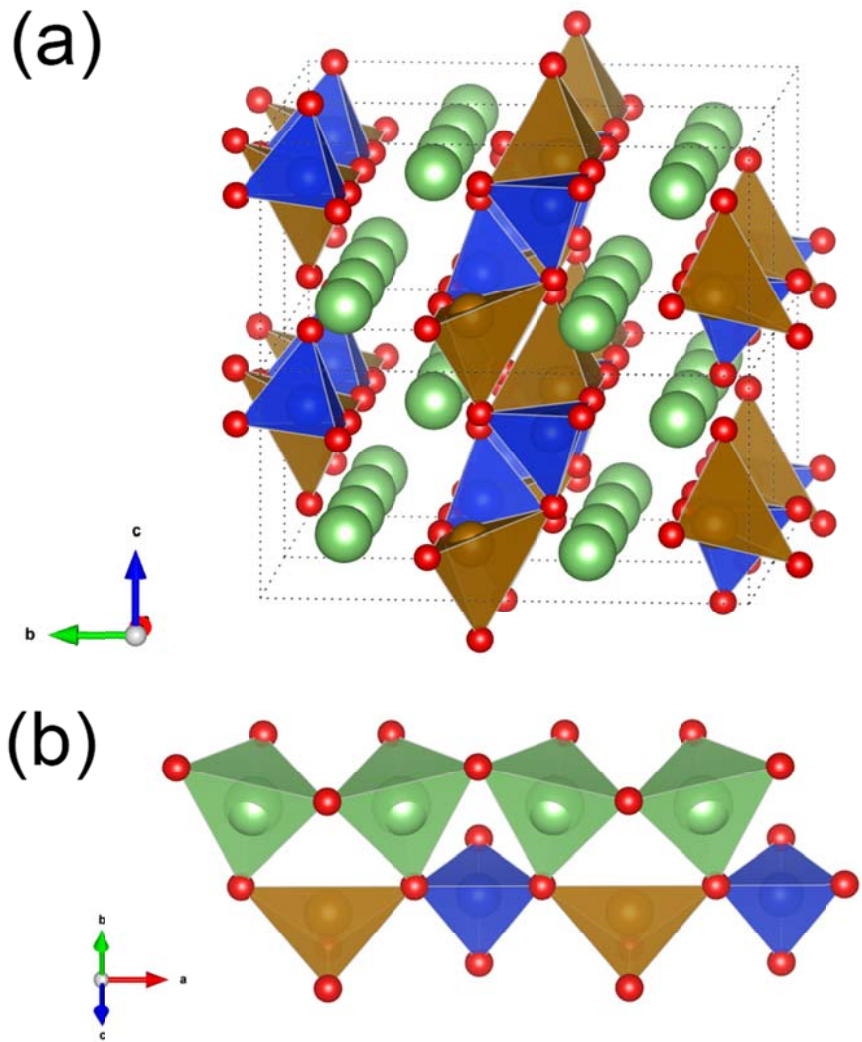

Figure 7: (a) Representation of the crystal structure of *Pmnb*-Li$_2$FeSiO$_4$ consisting of tetrahedra of FeO$_4$ (brown), SiO$_4$ (blue) and LiO$_4$ (green, tetrahedra removed for visibility). Oxygen ions are depicted in red. (b) Tetrahedral orientation along the *a* direction.

*4.2. Neutron diffraction*

The powder neutron diffractogram of $Li_2FeSiO_4$ shown in Fig. 8 confirms the *Pmnb*-structure and in addition enables the investigation of possible antisite defect formation in this material. In general, the crystal structure of lithium transition metal orthosilicates $Li_2MSiO_4$ (M = Mn, Fe) exhibits comparable Li – O and M – O bond lengths, which is known to give rise to Li – M antisite defects. Due to the large difference in the coherent neutron scattering lengths of Li and Fe (-1.90 and 9.45 fm, respectively) high-resolution neutron diffraction (ND) studies open the possibility to study the antisite disorder [3, 36, 37]. Generally, compounds with tetrahedrally coordinated ions are more prone to site exchange than compounds with octahedral coordination since the crystal field stabilization energies are usually lower [38]. Note, that DFT calculations on the *Pmnb*-polymorph suggest that additional exchange of Li and Fe over their primary sites is favourable upon delithiation [39]. The high-resolution ND pattern in Fig. 8 indicates cation mixing between the two crystallographic sites 8d and 4c. The data were recorded with a wavelength $\lambda$ = 1.051 Å in the paramagnetic regime at 25 K. Rietveld refinement in the *Pmnb* space group was done with total occupancy of each site constrained to unity. The analysis yields that the fit quality significantly improves when considering Li – Fe site exchange. Structural refinement was stable and the reliability factors minimized when assuming a fraction of around 2 % antisite defects for the final model.

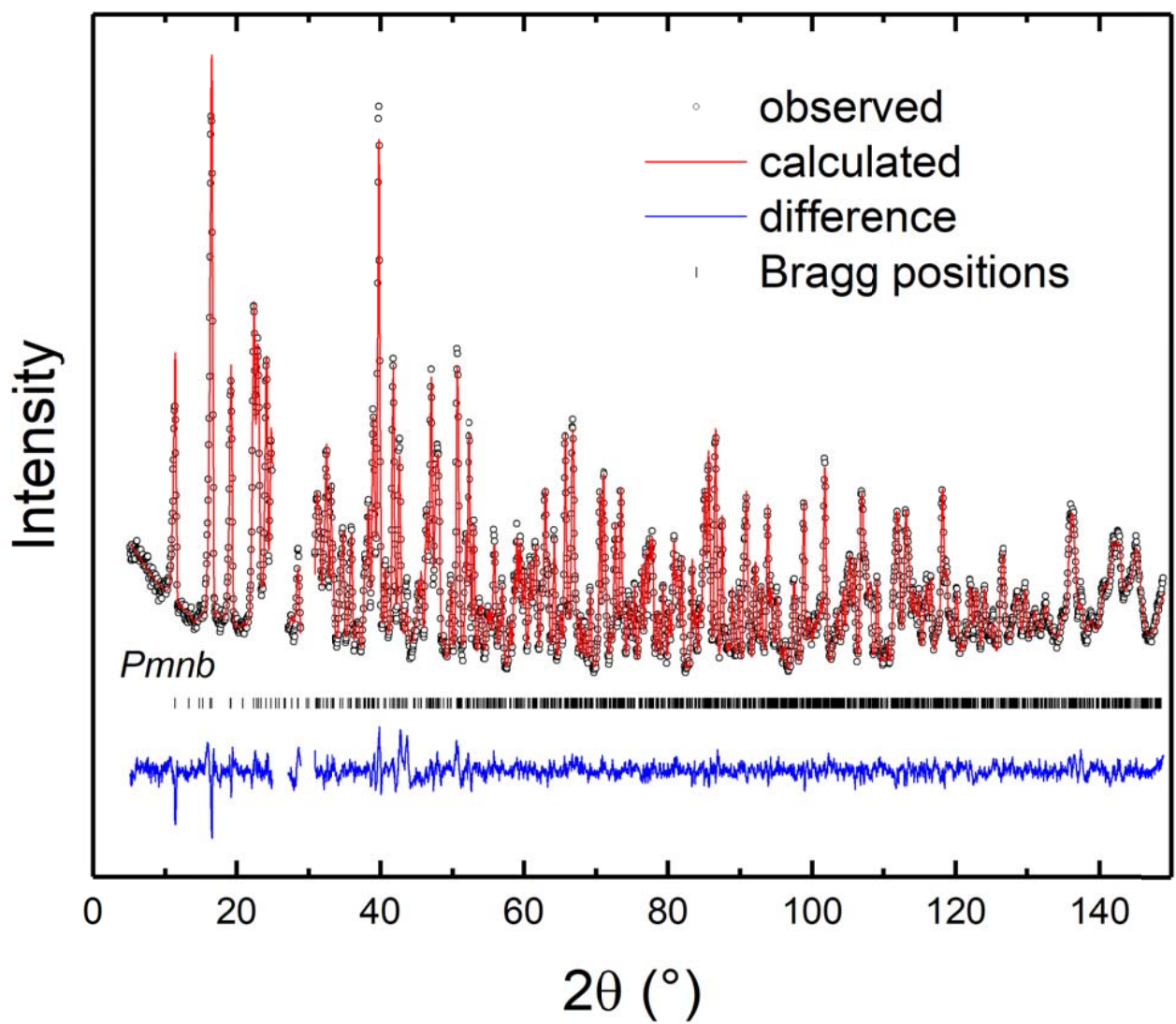

Figure 8: Observed (open circles), calculated (full line) and difference (bottom) intensities for the neutron Rietveld refinement of polycrystalline

$Pmnb$-Li$_2$FeSiO$_4$ at 25 K. Vertical ticks refer to the nuclear Bragg positions. Some regions with extra diffraction peaks due to the sample mount in a cryostat were excluded from the refinement. Reliability factors: $\chi^2$ = 2.57, Rwp = 9.25, Rp = 7.98, Bragg R-factor = 3.13.

*4.3. Magnetization and specific heat*

Studies on the anisotropic static susceptibility $\chi = M/B$ confirm the high quality of the single crystal. The data presented in Fig. 9 show Curie-Weiss-like behavior at high temperatures. For $B||a$, upon cooling there is a maximum at around $T_m$ = 28 K. Sharp decrease of $\chi$ below $T_m$ implies the onset of long-range antiferromagnetic (AFM) order with the crystallographic $a$-direction being the magnetic easy axis. For $B \perp a$, the onset of long-range antiferromagnetic order is associated with a small jump while there is a slight increase of $\chi$ below $T_N$. The magnetic specific heat $\partial(\chi_{mol}T)/\partial T$ (right inset in Fig. 9) exhibits a sharp $\lambda$-like anomaly. From the data, we read-off $T_N$ = 17.0(5) K, which is similar to previous studies on polycrystalline Li$_2$FeSiO$_4$ [2, 8, 40]. The upturn at lowest temperatures implies the presence of 0.5 % quasi-free spins, which is in full accordance to the $M$ vs. $B$ data (left inset of Fig. 9) and may be associated with a small amount of impurities. Note, that the signature of quasi-free moments is much weaker for $B||a$-axis, which indicates anisotropy. The anisotropy in $M$ vs. $T$ above room temperature is attributed to the anisotropy of the spectroscopic $g$-tensor in Li$_2$FeSiO$_4$. We note the absence of any signature of the Verwey transition, which excludes the presence of a Fe$_3$O$_4$ phase. If attributed to the Morin transition, a tiny feature in the magnetization data at $T \approx 275$ K, i.e., near to the bulk Morin temperature $T_M$ = 265 K would imply the presence of less than 0.03 % of $\alpha$-Fe$_2$O$_3$ [41]. Finally, we observe a broad and tiny anomaly around 220 K that may be attributed to a Fe$_{1-\delta}$O phase. Tentatively, comparing the magnetization data with Ref. [42] suggests an Fe$_{1-\delta}$O fraction of $\leq$ 0.5 %. The high crystal quality is further demonstrated by the sharp $\lambda$-shaped anomaly in the specific heat at the magnetic ordering temperature (Fig. 10). The data confirm the evolution of long-range magnetic order at $T_N$ as well as the continuous nature of the associated phase transition.

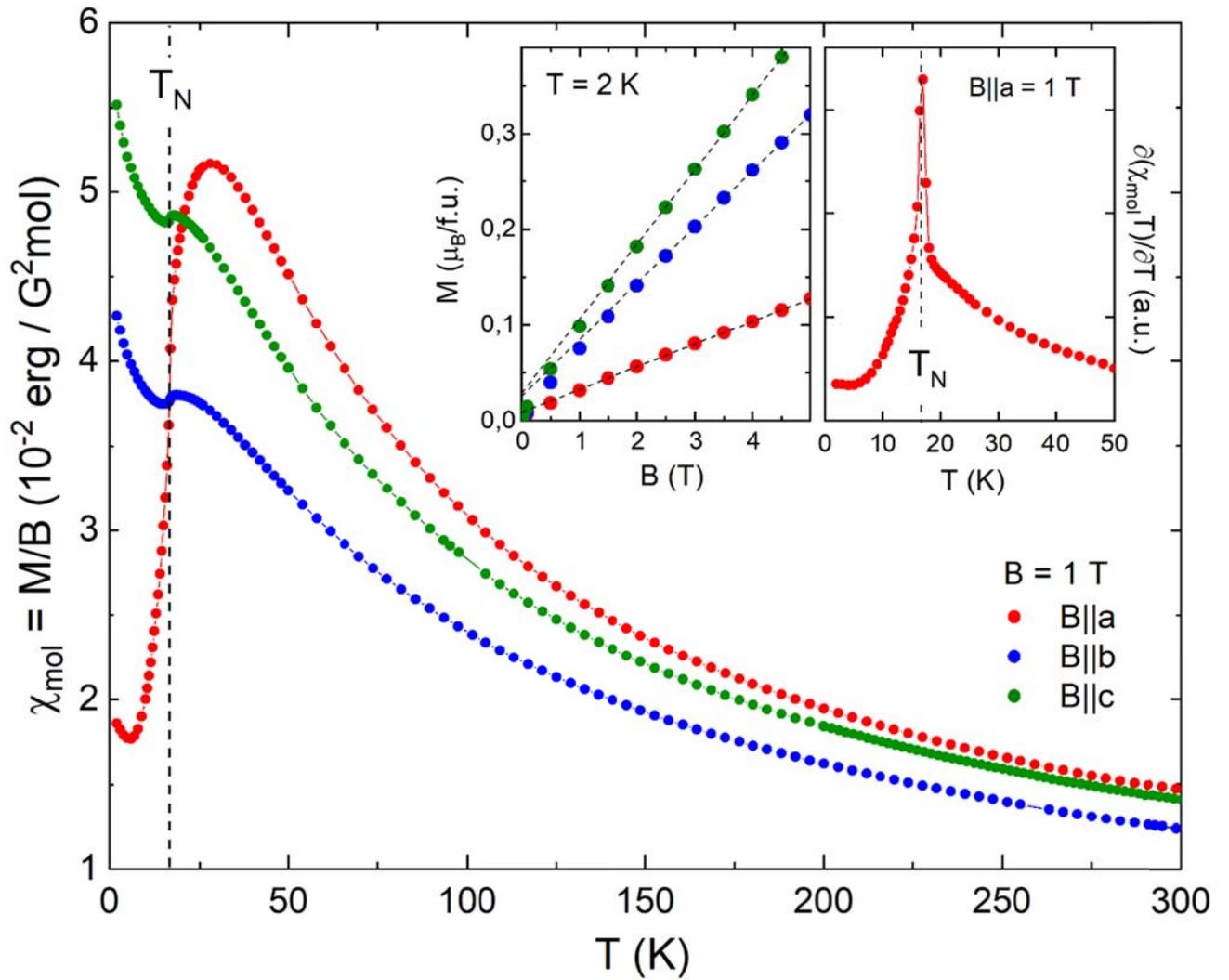

Figure 9: Temperature dependence of static magnetic susceptibility $\chi_{mol} = M/B$ up to $T = 300$ K measured in a magnetic field of $B = 1$ T applied along the principal crystallographic directions. The vertical dashed line marks the Néel temperature. The left inset depicts the field dependence of the magnetization, at $T = 2$ K. Dashed lines extrapolate $M(B > 4$ T$)$ to $B = 0$ T. The right inset shows magnetic specific heat $\partial(\chi_{mol}T)/\partial T$ obtained from measurement along the $a$-axis.

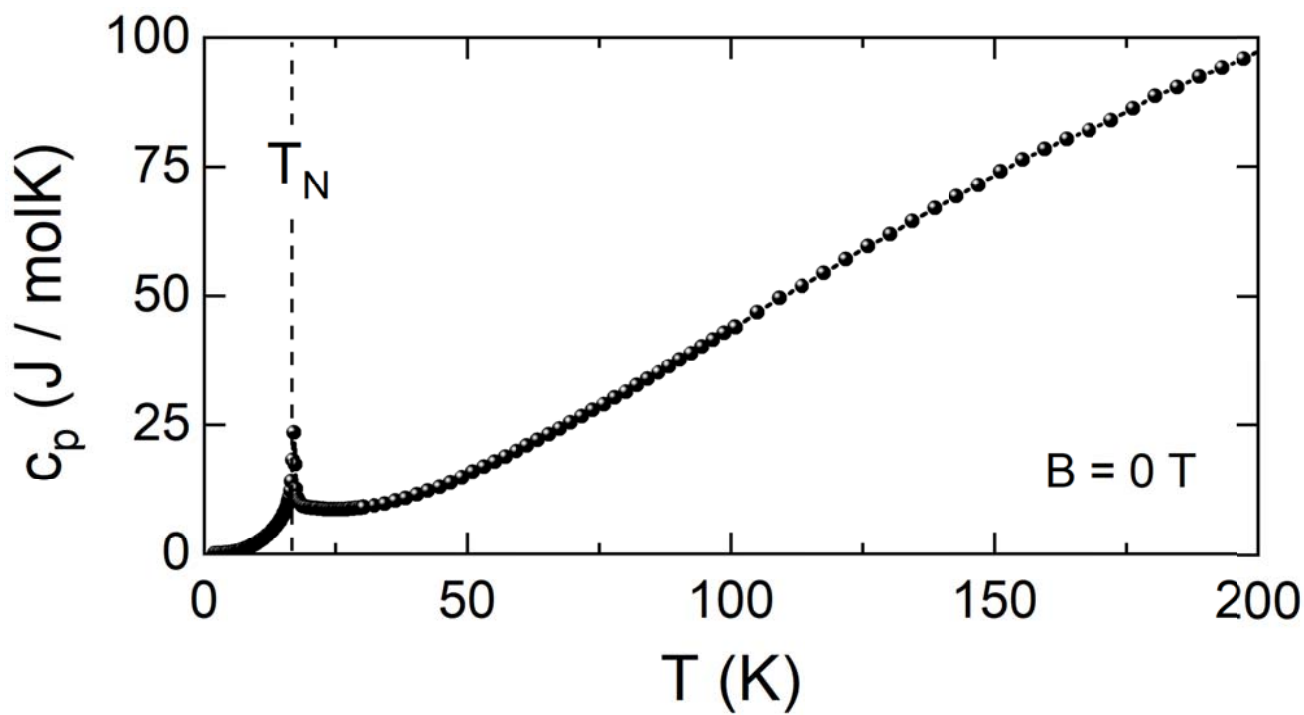

Figure 10: Temperature dependence of the specific heat $c_p$ in zero magnetic field. The vertical dashed line marks the Néel temperature.

## 5. Conclusions

Macroscopic single crystals of *Pmnb*-$Li_2FeSiO_4$ have been grown by the optical floating-zone technique under a high-pressure Ar atmosphere. The off-stoichiometric starting material exhibits $T_m$ = (1240 ± 5) °C. Employing slightly off-stoichiometric feed rods enables growth of mm-sized *Pmnb*-$Li_2FeSiO_4$ single crystals. The crystals were used to solve the crystal structure for the first time. The lattice constants were found to be $a$ = 6.27837(3) Å, $b$ = 10.62901(6) Å and $c$ = 5.03099(3) Å at 100 K. High-resolution neutron powder diffraction data suggest that a slight Li – Fe site exchange seems to be an intrinsic feature of *Pmnb*-$Li_2FeSiO_4$. The high quality of the grown single crystal is confirmed by sharp anomalies in the static magnetic susceptibility and in the specific heat associated with the onset of long-range antiferromagnetic order at $T_N$ = 17.0(5) K.

### 5.1. Acknowledgments

The authors thank Ilse Glass and Dr. Alexander Varychev for technical support. Financial support by the German-Egyptian Research Fund (GERF IV) through project 01DH17036 and by the Deutsche Forschungsgemeinschaft (DFG) through project KL1824/5 is gratefully acknowledged.

### 5.2. Appendix

CCDC 1859157 contains the supplementary crystallographic data for this paper. These data can be obtained free of charge from the Cambridge Crystallograhic Data Centre's and FIZ Karlsruhe's joint Access Service via https://www.ccdc.cam.ac.uk/structures/.